\documentclass[letterpaper, 10 pt, journal, twoside]{ieeetran}
\usepackage[neverdecrease]{paralist}

\usepackage{amsmath}
\usepackage{color}
\usepackage{tikz}
\usepackage{graphicx}
\usepackage{authblk}
\usepackage{mathtools}
\usepackage{amsfonts}
\usepackage{algorithm}
\usepackage[noend]{algpseudocode}
\usepackage{balance}
\usepackage{arydshln}

\usepackage[noadjust]{cite}

\newtheorem{problem}{\bf Problem}
\newtheorem{definition}{Definition}

\newtheorem{remark}{Remark}

\newtheorem{proposition}{Proposition}
\newtheorem{example}{Example}

\title{Provably Safe Cruise Control of Vehicular Platoons}

\author{Sadra Sadraddini, Sivaranjani S, Vijay Gupta and Calin Belta
\thanks{
Sadra Sadraddini and Calin Belta are with the Department of Mechanical Engineering, Boston University, Boston, MA. \{sadra,cbelta\}@bu.edu. Sivaranjani S and Vijay Gupta are with the Department of Electrical Engineering, University of Notre Dame, South Bend, IN. \{sseethar,vgupta2\}@nd.edu
\\ Calin Belta and Sadra Sadraddini were partially funded by NSF grants CPS-1446151 and CMMI-1400167. 
Vijay Gupta and Sivaranjani S were partially funded by the AFOSR grant FA 9550-15-1-0186 and NSF grants ECCS-1550016 and CNS 1239224. Sivaranjani S was partially supported by the Schlumberger Foundation Faculty for the Future Fellowship.
}
}

\begin{document}

\maketitle

\thispagestyle{empty}
\pagestyle{empty}

\begin{abstract}
We synthesize performance-aware safe cruise control policies for longitudinal motion of platoons of autonomous vehicles. Using set-invariance theories, we guarantee infinite-time collision avoidance in the presence of bounded additive disturbances, while ensuring that the length and the cruise speed of the platoon are bounded within specified ranges. We propose (i) a centralized control policy, and (ii) a distributed control policy, where each vehicle's control decision depends solely on its relative kinematics with respect to the platoon leader. Numerical examples are included. 
 
\end{abstract}

% Keywords appear just beneath the abstract. Use only for final version.

\section{Introduction}

\IEEEPARstart{P}latooning
 strings of autonomous vehicles equipped with sensing, communication and computation capabilities has two promising benefits. First, there is a potential to dramatically increase traffic flow capacity since autonomous vehicles can move with small separations at high velocities \cite{varaiya1993smart}, a behavior that cannot be safely executed by human drivers. Second, platooning with tight spacing impacts the aerodynamics of vehicles in a manner that decreases fuel consumption \cite{al2010experimental}. 

Control-theoretic aspects of platoons have been studied for several decades (see \cite{jovanovic2015vehicular} and \cite{li2015overview} \color{black}  for a survey). The typical control objective has been asymptotic string stability, that is, the convergence of inter-vehicular spacings to desired set-points, specified in terms of constant spacing \cite{caudill1977vehicle,swaroop1999constant}\nocite{hedrick1994control,sheikholeslam1992control} or a constant time headway \cite{ioannou1993autonomous,li2002traffic}. String stability has been studied for various information architectures such as predecessor following \cite{peppard1974platoon,stankovic2000distributed}, leader and predecessor following \cite{sheikholeslam1992control,sabuau2017optimal}, and nearest neighbor interaction \cite{seiler2001coordinated,barooah2009mistuning,zhang1999using}. $H_\infty$ methods have also been proposed to deal with exogenous disturbances \cite{seiler2001coordinated,zheng2016platooning}.  Recently, multi-agent control paradigms have been employed to analyze vehicular platoons from a scalability perspective for various information topologies \cite{zheng2016scalability} and design distributed controllers to enhance stability margins in large platoons \cite{zheng2016stability}.

While asymptotic string stability is important to guarantee that small disturbances do not propagate along the platoon, it is more important to guarantee the safety of the platoon in the sense that vehicles do not collide with each other. Early works on safe longitudinal platooning \cite{lygeros1998verified,alvarez1999safe,tomlin2000game} focused on simple two-car scenarios where the input to the leader vehicle is considered as an adversarial input and the objective of the follower vehicle is to avoid collision with the leader for all allowable behaviors of the leader.  Such a formalism may be too conservative since it does not assume intelligent control over the leader. A similar paradigm was considered in \cite{alam2014guaranteeing,turri2017cooperative} for platooning of heavy duty trucks, where the platoon is considered safe, if for all possible behaviors of a vehicle in the platoon, there exist control inputs for other vehicles to avoid collisions. A control invariant set was synthesized to restrict the evolution of the system accordingly.

Instead of considering driver behavior of the predecessor vehicle as adversarial, a more realistic and less conservative approach is to consider platooning when fast, time varying disturbances act in an adversarial manner on the vehicles. Disturbances are physically attributed to hard-to-model nonlinear  effects such as changes of gears, side wind effects and actuator imperfections. Further, adversarial disturbances may also be used, to an extent, to capture the effect of communication delays and other delays in the platoon dynamics. The objective, then, is to design controllers such that collisions are avoided for all times and all allowable disturbances. It is well known that such a formulation is closely related to $l_\infty$ optimal control \cite{vidyasagar1986optimal}, viability theory \cite{aubin2009viability} and set-invariance methods \cite{Blanchini:1999aa}. 
%Invariance inducing controllers are often nonlinear, even for linear systems.
Controllers satisfying set-invariance specifications are often nonlinear even if the system is linear \cite{kerrigan2001}, while no invariance inducing linear controller may exist. Therefore, stabilizing linear controllers may not be sufficient for platoon safety.
%As invariance inducing linear controllers may not exist even for linear systems when a nonlinear one exists \cite{kerrigan2001}, stabilizing linear controllers may not be sufficient for platoon safety. 
While recent works like \cite{huang2016toward}, \cite{verginis2017robust} guarantee safety by designing controllers to bound the spacing error of vehicles, these methods do not provide specific bounds on the control inputs.  Therefore,  exhaustive simulations and tuning of parameters are required to verify whether the platoon operation is collision-safe and %control 
	actuator limits are respected.
 
{We provide a formal correct-by-design approach to synthesize platoon safety control policies using set-invariance methods}. We model each vehicle as a discrete-time double integrator subject to bounded additive disturbances and bounded control inputs. We specify the system requirements as infinite-time collision avoidance while restricting the length and the cruise speed of the platoon to user-defined ranges. We propose both centralized and distributed optimal control policies. For the centralized case, we compute a robust control invariant (RCI) set inside the safe set corresponding to the platoon specifications. We use the computational methods from \cite{rakovic2007optimized} to synthesize RCI sets by solving a single linear program, which has polynomial complexity in the size of platoon. However, this method is still not scalable to large platoons. Therefore, using ideas from separable RCI sets \cite{Nilsson2016seperable}, we propose a (conservative) distributed control policy with $\mathcal{O}(1)$ complexity, where each vehicle's control decision depends solely on its relative kinematics with respect to the platoon leader. 

This paper is organized as follows. In Sec. \ref{sec:problem}, we formulate the problem. In Sec. \ref{sec:solution}, we use set invariance theories to design centralized and distributed controllers that solve the safety problem, and provide numerical simulations.

\section*{Notation}
We denote the sets of real and non-negative real numbers by $\mathbb{R}$ and $\mathbb{R}_+$, respectively. We denote the vector of all ones $(1,\cdots,1)^T \in \mathbb{R}^n$ by $1_n$ and the identity matrix in $\mathbb{R}^{n\times n}$ by $I_n$. Given a set $X$, its cardinality is denoted by $|X|$. For $X \subset \mathbb{R}^n$, the projection of $X$ on the $i$-th Cartesian direction is denoted by $Proj_i X$. Given sets $X,Y \subset \mathbb{R}^n$, their Minkowski sum is represented by $X \oplus Y$. Given a matrix $A \in \mathbb{R}^{n\times n}$ and $X \subset \mathbb{R}^n$, $AX$ is interpreted as $\{Ax | x\in X\}$.  Given sets $X_i$, $i=1,\cdots,n$, we use $\prod_{i=1}^n X_i$ to denote Cartesian product $X_1 \times X_2 \times \cdots X_n$. Given functions $f_i:X_i \rightarrow Y_i$, $i=1,\cdots,n$, we denote the product function $f:=<f_i>_{i=1,\cdots,n}$ as $f:\prod_{i=1}^n X_i \rightarrow \prod_{i=1}^n Y_i$, $f(x_1,\cdots,x_n)=(y_1,\cdots,y_n)$ such that $y_i=f(x_i), x_i \in X_i, y_i \in Y_i, i=1,\cdots,n$.  

\section{Problem Statement and Approach}
\label{sec:problem}
\begin{figure}[t]
\centering
\vspace{2pt}
\begin{tikzpicture}[xscale=2.10,yscale=0.7]
    \node[] at (0,0) {\includegraphics[width=0.1\textwidth]{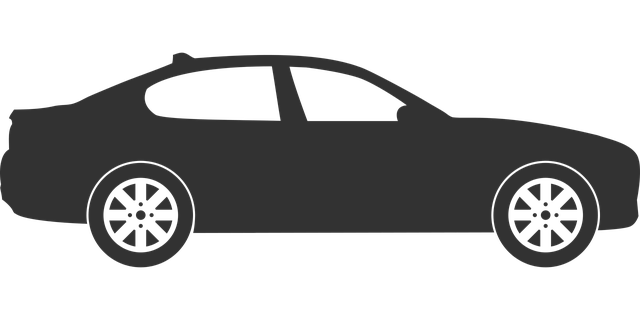}};
    \node[] at (1,0) {\includegraphics[width=0.1\textwidth]{car}};
    \node[] at (2,0) {\includegraphics[width=0.1\textwidth]{car}};
    \node[] at (3,0) {\includegraphics[width=0.1\textwidth]{car}};
    \draw[color=black,<-]  (3.5, -0.42 ) -- (-0.46, -0.42 );
    \draw[color=black!50,dashed,thin]  (3.42, 1.6 ) -- (3.42, -0.2 ); % Car 0
    \draw[color=black!50,dashed,thin]  (2.41, 0.6 ) -- (2.41, -0.2 ); % Car 1
    \draw[color=black!50,dashed,thin]  (1.42, 1.1 ) -- (1.42, -0.2 ); % Car 2
    \draw[color=black!50,dashed,thin]  (0.41, 1.6 ) -- (0.41, -0.2 ); % Car 3
    
    \draw[color=black,->]  (3.42, 0.6 ) -- (2.41, 0.6 );
    \draw[color=black,->]  (3.42, 1.2 ) -- (1.42, 1.2 );
    \draw[color=black,->]  (3.42, 1.8 ) -- (0.41, 1.8 );
    
    \node[] at (2.95,0.9) {$\tilde{x}_1$};
    \node[] at (2.45,1.5) {$\tilde{x}_2$};
    \node[] at (1.95,2.1) {$\tilde{x}_3$};
    
    \node[] at (3,-0.7) {$0$};
    \node[] at (2,-0.7) {$1$};
    \node[] at (1,-0.7) {$2$};
    \node[] at (-0.0,-0.7) {$3$};
\end{tikzpicture}
\caption{A platoon of 4 vehicles}
%\vspace{-1.5em}
\label{fig:platoon}
\end{figure}

Consider a platoon of $N+1$ autonomous vehicles moving on a single lane road. The index of the leader vehicle is designated to be $0$ and the indices of the follower vehicles are in increasing order (see Fig. \ref{fig:platoon}). The discrete-time dynamics of each vehicle $i \in \{0,1,\cdots,N\}$ is modeled as:
\vspace{-0.05em}
\begin{equation}
\begin{array}{l}
\label{eq:absolute}
{x}^+_i=x_i+v_i \Delta \tau + u_i \frac{\Delta \tau^2}{2} + w_{i,x}, \\
{v}^+_i=v_i+u_i \Delta \tau +w_{i,v}, \\
\end{array}
%\vspace{-0.45em}
\end{equation}
where $x_i$, $v_i$ and $u_i$ are the absolute (in the ground (road) frame) position (measured as the front of the vehicle), velocity and control input respectively, $w_{i,x}$ and $w_{i,v}$ are the additive disturbances affecting the position and the velocity respectively, and $\Delta \tau$ is the sampling time. The \emph{platoon velocity} is defined as the velocity of the leader, denoted by $v_0$, and the \emph{platoon length} is defined as the difference between the positions of the leader and the $N$-th vehicle, $x_0-x_N$. 
We denote the length of the $i$'th vehicle by $l_i \in \mathbb{R}_+$. The platoon length is  physically lower-bounded by $\sum_{i=0}^{N} l_i$.
The control input of the platoon is defined as $u:=\{u_i\}_{i=0,1,\cdots,N}, u \in \mathbb{U}$, where each vehicle has a bounded range of admissible control inputs: 
\begin{equation}
\label{eq:u}
\mathbb{U}:=\prod_{i=0}^N [u_{i,\min},u_{i,\max}].
\end{equation}
Similarly, we define the platoon disturbance vector as $w:=\{w_{i,s}\}_{s=x,v; i=0,1,\cdots,N}, w \in \mathbb{W}$, where the admissible range of disturbances on each vehicle is bounded: 
\begin{equation}
\label{eq:w}
\mathbb{W}:=\prod_{i=0}^N [w_{i,x,\min},w_{i,x,\max}] \times [w_{i,v,\min},w_{i,v,\max}].
\end{equation}
Note that $\mathbb{U} \subset \mathbb{R}^{N+1}$ and $\mathbb{W} \subset \mathbb{R}^{2(N+1)}$, and both are hyper-rectangles in their respective domains. The bounds in $\mathbb{W}$ may depend on the vehicle velocities, leading to nonlinearities in \eqref{eq:absolute}. For simplicity, we assume that the bounds in $\mathbb{W}$ are constant and known given a nominal operating point of \eqref{eq:absolute}.  
While the determination of these bounds is out of scope of this paper, a suitable experimental setup can be used to compute a (tight) hyper-rectangle $\mathbb{W}$ containing all measured disturbances. {The exact knowledge of $\mathbb{W}$ is not required. As it will be clear from subsequent discussion, over-approximating $\mathbb{W}$ retains collision avoidance guarantees, but is conservative. Under-approximation, on the other hand, may lead to collision.}

We assume that the vehicles do not have (accurate, real-time) access to their absolute positions, but have real-time access to relative distances using on-board radars and relative velocities. Vehicles can measure their absolute velocities using on-board speedometers. Relative velocities can then be computed from data acquired by Vehicle-to-Vehicle (V2V) communication. We define the relative position and velocity of the $i$-th vehicle with respect to the leader to be
%\vspace{-0.45em}
\begin{equation}
\label{eq:relative}
\tilde x_i:=x_0-x_i, \tilde v_i:=v_0-v_i, i=1,\cdots,N.
\vspace{-0.45em}
\end{equation} 
Substituting \eqref{eq:relative} in \eqref{eq:absolute}, we obtain:
%\vspace{-0.45em}
\begin{equation}
\small{
\label{eq:relative_dynamics}
\begin{array}{l}
\tilde{x}^+_i=\tilde x_i+\tilde v_i \Delta \tau + (u_0-u_i) \frac{\Delta \tau^2}{2} + w_{0,x}-w_{i,x}, \\
\tilde {v}^+_i=\tilde v_i+(u_0-u_i) \Delta \tau +w_{0,v}-w_{i,v}, i=1,\cdots,N, \\
v_0^+= v_0+ u_0 \Delta \tau +w_{0,v}. 
\end{array}}
%\vspace{-0.45em}
\end{equation}
       
The \emph{platoon state} is defined as the vector $y:=(\tilde x_1,\tilde v_1,\tilde x_2,\tilde v_2,\cdots, \tilde x_N,\tilde v_N,v_0)^T$, $y\in \mathbb{R}^{2N+1}$. The evolution of the platoon state is given by: 
%\vspace{-0.45em}
\begin{equation}
\label{eq:system}
y^+=Ay+Bu+Ew,
%\vspace{-0.45em}
\end{equation}
where $A$, $B$ and $E$ are constant matrices obtained from \eqref{eq:relative_dynamics}.

%\vspace{-0em}

\begin{remark}[Information Structure]
	The requirement that all vehicles possess information about the leader's position and velocity may lead to performance issues in large platoons due to communication delays. However, taking into account imperfections through disturbances can potentially allow the use of larger sampling times $\Delta t$, thereby accommodating greater communication times and hence, longer platoons. Future work will involve relaxing the V2V communication demand on the platoon by considering local information from preceding and following vehicles.
\end{remark}
We now state the problem addressed in this paper.
\begin{problem}
\label{problem:problem}
Given a platoon of $N+1$ autonomous vehicles whose dynamics is described by \eqref{eq:system}, a desired upper-bound for platoon length $L>\sum_{i=0}^{N} l_i$ and a desired range of platoon speed $[v_{0,\min},v_{0,\max}]$, find a feedback control policy $\mu:\mathbb{R}^{2N+1} \rightarrow \mathbb{U}, u=\mu(y),$ such that:
\begin{itemize}
\item collisions are avoided\footnote{Theoretically, collision-avoidance inequality constraints are strict. However, our computational methods (e.g., linear programming) can only handle non-strict inequalities. {In practice, due to over-approximations of the disturbances, inequality constraints are satisfied with some margin and non-strict formulation does not raise a problem.}}: 
\begin{equation}
\label{eq:collide}
\tilde x_i > \tilde x_{i-1} + l_{i-1},~i=1,\cdots,N,~\tilde x_0=0,
\end{equation}
\item the platoon length is bounded: 
\begin{equation}
\label{eq:length}
\tilde x_N \le L, \text{\space and},
\end{equation}
\item the platoon velocity is bounded: 
\begin{equation}
\label{eq:speed}
v_0 \in [v_{0,\min},v_{0,\max}],
\end{equation}
\end{itemize}
for all times and for all allowable sequences of admissible disturbances acting on the platoon. Denote by $\mathbb{M}$ the set of all control policies such that \eqref{eq:collide}-\eqref{eq:speed} are satisfied. If $|\mathbb{M}|>1$, select an optimal control policy $\mu^*$ such that $J(\mu^*) \le J(\mu), \forall \mu \in \mathbb{M}$, where $J: \mathbb{M} \rightarrow \mathbb{R}$ is a cost function. 
\end{problem}
\vspace{5pt}

Note that we have not specified desired speed ranges for the follower vehicles, as these naturally follow from the bounds on the platoon length and the collision avoidance requirements. Nevertheless, speed specifications for individual vehicles can be easily accommodated in our framework. Constraints \eqref{eq:collide}-\eqref{eq:speed} define a convex safe set $\mathbb{S} \subset \mathbb{R}^{2N+1}$. The problem of finding a safe control policy is equivalent to finding a robust control invariant set inside $\mathbb{S}$, which is formalized in Section \ref{sec:invariance}.

\section{Safe Cruise Control}
\label{sec:solution}
We provide two solutions to Problem \ref{problem:problem}. First, we propose a centralized solution where $u=\mu(y)$ is computed by a central coordinator (which can be placed on one of the vehicles, say the leader). Even though the computation of this solution has polynomial complexity in the size of platoon, it is not applicable to very large platoons. Second, we provide a distributed solution in which the control decision for each of the following vehicles is computed solely using their kinematics with respect to the leader.  

\subsection{Controlled Invariance}
\label{sec:invariance}
\begin{definition}
A set $\Omega \subset \mathbb{R}^{2N+1}$ is a \emph{robust control invariant} (RCI) set for system \eqref{eq:system} if $$\forall y \in \Omega, \exists u \in \mathbb{U}, \text{ s.t. } Ay+Bu \oplus E\mathbb{W} \subseteq \Omega.$$
\end{definition}
Given a RCI set $\Omega \subseteq \mathbb{S}$, it is well known that there exists a unique, maximal RCI $\Omega_{\infty} \subseteq \mathbb{S}$ which contains all the RCI sets in $\mathbb{S}$.  The computation of $\Omega_{\infty}$ requires implementing the well known fixed-point algorithm \cite{kerrigan2001}:
%\vspace{-0.45em}
\begin{equation}
\label{eq:fixed}
\Omega_0=\mathbb{S}, \;
\Omega_{k+1}=\Omega_k \cap \text{rPre}(\Omega_k),
%\vspace{-0.45em}
\end{equation} 
where $\text{rPre}(\Omega_k)=\{y \in \mathbb{R}^{2N+1} | \exists u \in \mathbb{U},\text{ s.t. } Ay+Bu\oplus E\mathbb{W} \subseteq \Omega_k \}$. We have $\Omega_k \subseteq \Omega_{k-1} \subseteq \cdots \subseteq \Omega_0=\mathbb{S}$. The algorithm terminates at $k^* \in \mathbb{N}$ if $\Omega_{k^*}=\Omega_{k^*-1}$, which concludes $\Omega_{\infty}=\Omega_{k^*}$. Then, we have:
\begin{equation*}
\mathbb{M}:=\left\{ \mu:\Omega_\infty \rightarrow \mathbb{U} \big| Ay+B\mu(y) \oplus E\mathbb{W} \subseteq \Omega_\infty \right\}.
\end{equation*}
Notice that $\mathbb{M}=\emptyset$ if and only if $\Omega_{\infty}=\emptyset$. There is no guarantee that there exists a finite $k^*$ to terminate \eqref{eq:fixed}. Therefore, determining a safe control policy belongs to the \emph{semi-decidable} class of decision problems in the sense that the corresponding algorithm eventually terminates if a safe control policy exists, but may run indefinitely if one does not exist. In addition to the termination issue, implementing \eqref{eq:fixed} is computationally difficult even for small dimensions due to the quantifier elimination process and numerical issues that arise from handling a large number of constraints with many potential redundancies \cite{kerrigan2001}. Therefore, we follow a different approach to find a RCI set for \eqref{eq:system}.

The idea is to find a RCI set $\Omega \subseteq \Omega_\infty \subseteq \mathbb{S}$ instead of the maximal RCI set. Therefore, safety is guaranteed (soundness) but completeness may be lost. There exist methods that (under-)approximate the maximal RCI set \cite{Blanchini:1999aa}, \cite{rungger2017computing}. 
We use the method introduced by Rakovi{\'c} et al. \cite{rakovic2007optimized} to compute a RCI set. 
{ 
	{Due to space limitations, we briefly explain the method to find a RCI set} for a linear system characterized by the pair $(A \in \mathbb{R}^{n\times n},B \in \mathbb{R}^{m\times n})$ and the additive disturbance set $\mathbb{D}=E\mathbb{W} \subset \mathbb{R}^n$. For system \eqref{eq:system}, we have $n=2N+1$ and $m=N+1$.  
\begin{figure*}[t]
\centering
\vspace{1pt}
\begin{tikzpicture}[xscale=.5,yscale=0.4]
    \node[] at (0,0) {\includegraphics[width=0.23\textwidth]{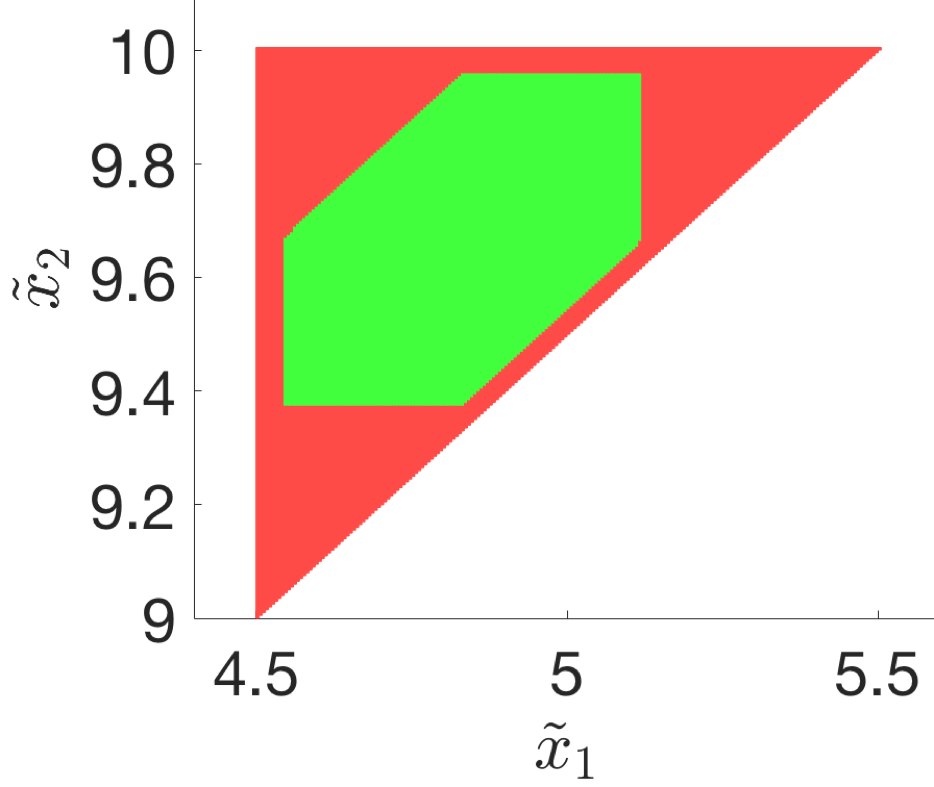}};
    \node[black] at (-0.1,1.5) {\Large $\Omega^*$};
    \node[black] at (-1.4,3) {\Large $\mathbb{S}$};
    \node[] at (9,1.5) {\includegraphics[width=0.18\textwidth]{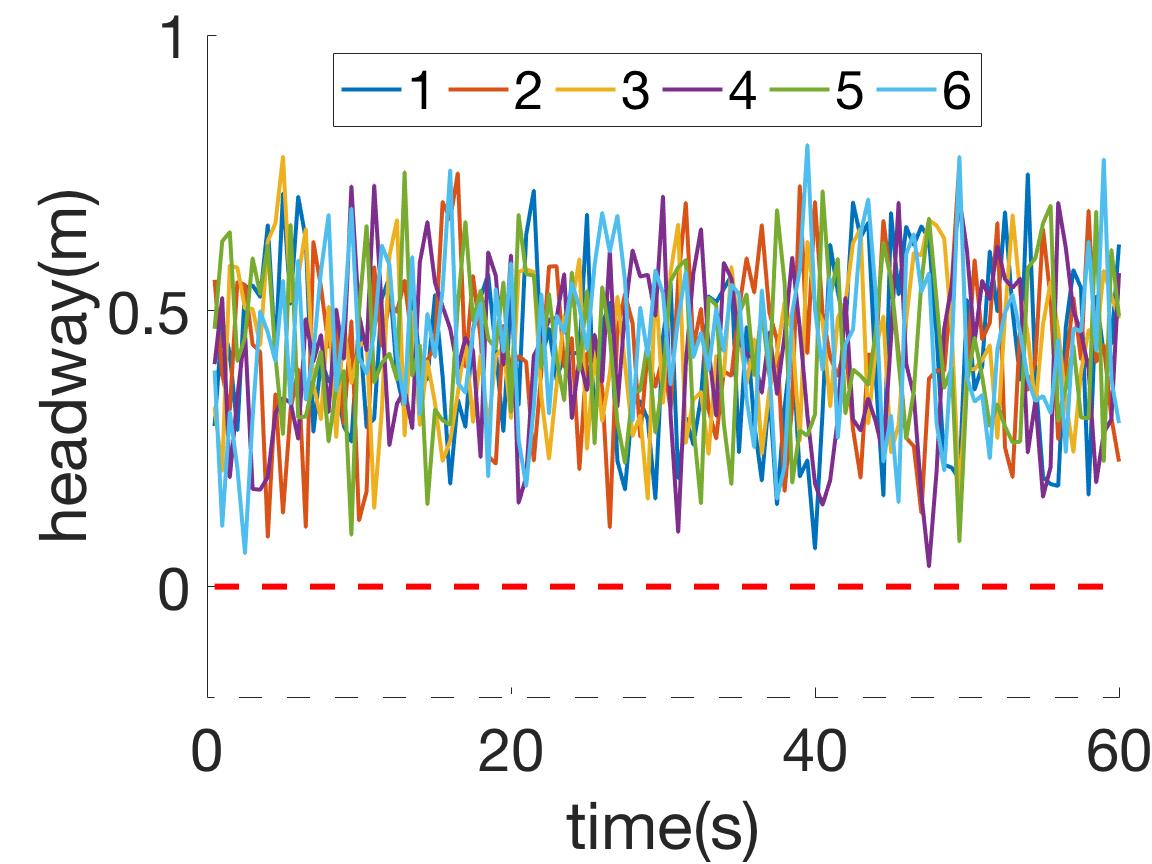}};
	\node[] at (17,1.5) {\includegraphics[width=0.18\textwidth]{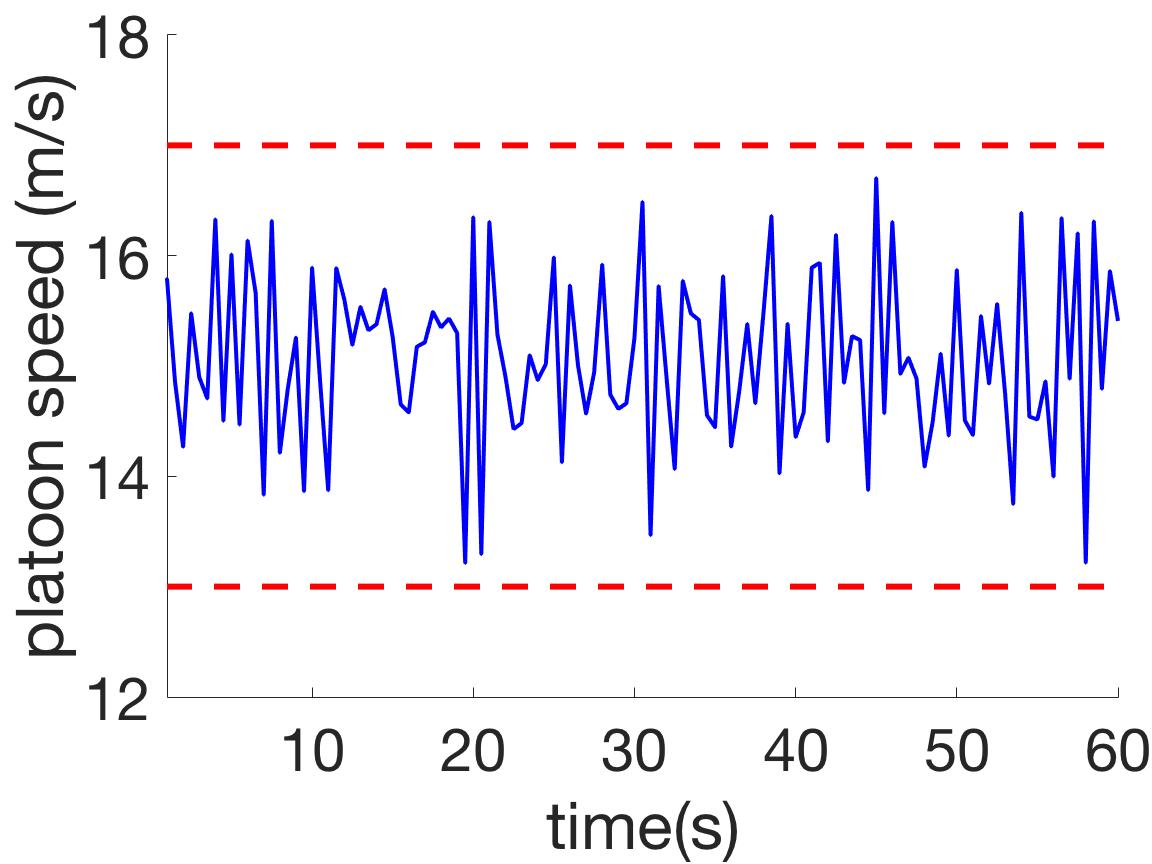}};
	\node[] at (25,1.5) {\includegraphics[width=0.18\textwidth]{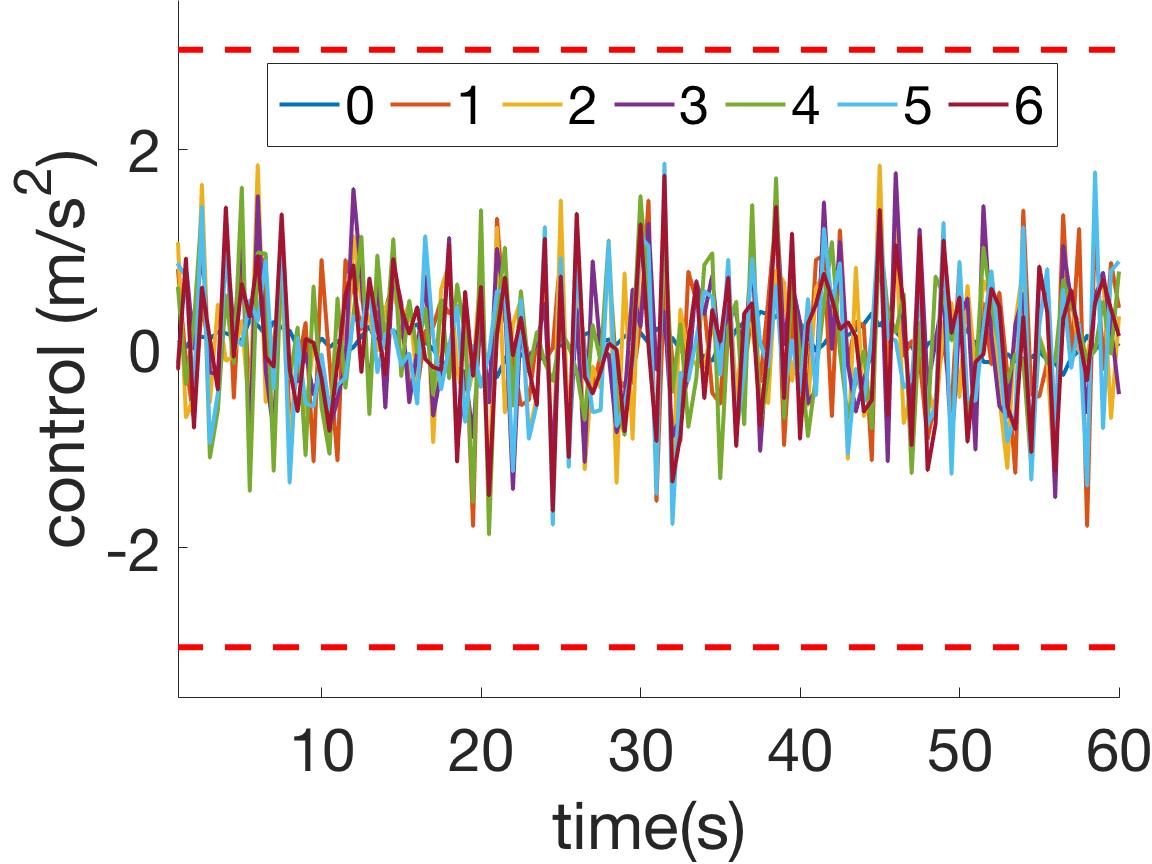}};
	\node[] at (17,-3.3) {\includegraphics[width=0.65\textwidth]{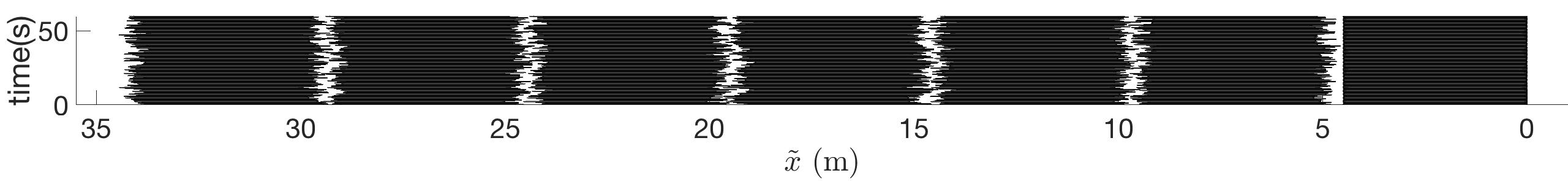}};
\end{tikzpicture}
\vspace{-0.0em}
\caption{Example 1: [left] Projection of the safe set $\mathbb{S}$ and the RCI set $\Omega$ on $\tilde x_1$-$\tilde x_2$ space. [right] Simulation of platoon of $N=6$. [Top, left to right]: Trajectories of headway, platoon speed and controls versus time [Bottom] Vehicular displacements in the leader frame.}
\label{fig:invariant}
%\vspace{-1em}
\end{figure*}
Consider the following set:
%\vspace{-0.45em}
\begin{equation}
\Omega({\bf M}_\kappa)=\bar{x} \oplus (1-\alpha)^{-1} \bigoplus_{i=0}^{\kappa-1} (A^i+\xi_i {\bf M}_\kappa)\mathbb{D},
%\vspace{-0.45em}
\end{equation}
where $\bar{x} \in \mathbb{R}^n$ is an offset vector, ${\bf M}_\kappa=(M_0^T,M_1^T,\cdots,M_{\kappa-1})^T \in \mathbb{R}^{\kappa m \times n}$ is the matrix of parameters and $\xi_i=(A^{i-1}B, A^{i-2}B, \cdots, AB, B, 0 , \cdots, 0) \in \mathbb{R}^{n \times \kappa m},~i=0,\cdots,\kappa,~\xi_0=0^{n \times \kappa m}$, where $\kappa$ is an integer,  %which has to be 
{ greater than the controllability index of the pair $(A,B)$, that determines the number of parameters} characterizing the RCI set, and $\alpha \in [0,1]$ is the contraction factor.   
\begin{proposition}\cite{rakovic2007optimized}
The set $\Omega(M_\kappa)$ is a RCI set in $\mathbb{S}$ if there exists ${\bf M}_\kappa$ such that
\begin{subequations}
\label{eq:rci}
%\vspace{-0.45em}
\begin{equation}
(A^\kappa+\xi_\kappa {\bf M}_\kappa) \mathbb{D} \subseteq \alpha \mathbb{D},
\end{equation}
%\vspace{-0.95em}
\begin{equation}
\bar{u} \oplus (1-\alpha)^{-1} \bigoplus_{i=0}^{\kappa-1} ({\bf M}_i\mathbb{D}) \subseteq \mathbb{U},
\end{equation}
\end{subequations}
where $\bar{u}$ is an offset vector for the controls. 
\end{proposition}
It can be shown that checking the feasibility of the constraints in \eqref{eq:rci}, alongside $\Omega({\bf M}_\kappa) \subseteq \mathbb{S}$, is equivalent to feasibility checking of a linear program (LP). Moreover, by solving the corresponding LP with a zero cost vector or some ad-hoc cost criteria\footnote{See \cite{rakovic2007optimized} for some useful cost vector suggestions.},  numerical values for $\bar{x}, \bar{u}$ and  $\bold{M}_\kappa$ that characterize the RCI set are found\footnote{Our software implementing the method in  \cite{rakovic2007optimized} is publicly available at \texttt{https://github.com/sadraddini/PARCIS}}. 
In many cases, the RCI sets computed using this method finely under-approximate, or are exactly equal to, the maximal RCI set. However, this method has the disadvantage of not providing an explicit representation for the RCI set, instead providing an implicit representation in high dimensions. Nevertheless, the invariance-inducing controller can be computed in high dimensions and mapped back to the original dimensions. The details of this method are not provided here as they are well documented in \cite{rakovic2007optimized}.   
}

In this paper, the admissible disturbance set $\mathbb{W}$ plays a major role in determining whether a safe control policy exists. We let 
\begin{equation}
\mathbb{W}=\lambda \mathbb{W}_{0},
\end{equation}
where $\mathbb{W}_{0} \subseteq \mathbb{R}^{2(N+1)}$ is a fixed set (assumed to be a hyper-rectangle, as previously stated), and $\lambda>0$ is a scalar that is used to scale the set. Using the bisection method over $\mathbb{R}_+$, we search for the largest $\lambda$, denoted by $\lambda^*$, such that a RCI set exists. In this context, $\lambda^* \mathbb{W}_{0}$ is the largest set of admissible disturbances that the platoon can accommodate. We denote the corresponding RCI set by $\Omega^*$. 

\subsection{Centralized Control}
\label{sec:centralized}
We compute a RCI set $\Omega \subseteq \mathbb{S}$. Define
\begin{equation*}
\mathcal{M}:=\left\{ \mu:\Omega \rightarrow \mathbb{U} \big| Ay+B\mu(y)+E\mathbb{W} \subseteq \Omega \right\}.
\end{equation*}
%The following statement follows from the definition of RCI sets.    
For all $\mu \in \mathcal{M}$, constraints 
\eqref{eq:collide}-\eqref{eq:speed} are satisfied for all times if $y_0 \in \Omega$, where $y_0$ is the initial state. 
We select the suboptimal invariance-inducing control policy $\mu^*$ such that  $J(\mu^*) \le J(\mu), \forall \mu \in \mathcal{M}$ (global optimality is achieved if $\mathcal{M}=\mathbb{M}$, or $\Omega=\Omega_\infty$). It can be shown that the resulting control policy is a piecewise affine function of state, which can be computed explicitly in an offline manner using parametric programming \cite{kvasnica2004multi} or solved online using a linear/quadratic program. This control policy is centralized and the control decision is computed using the full knowledge of the state of the system.

\begin{example}
\label{example:centralized}
Consider a platoon of $N+1$ vehicles. Let $l_i=4.5\text{m}, i=1,\cdots,N,$ and the desired platoon speed range be $[13,17] \text{m/s}$. We assume $\Delta\tau=0.5\text{s}$, $\mathbb{U}=\prod_{i=0}^N [-3,3] \text{m/s}^2$, and $\mathbb{W}_0=\prod_{i=0}^N [-1,1]\text{m/s} \times [-0.25,0.25]\text{m}$. We use the method in \cite{rakovic2007optimized} with $\alpha=0$ and $\kappa=10$.   
\setdefaultleftmargin{0pt}{}{}{}{}{}
\begin{enumerate}
\item Let $N=2$ and let the upper-bound for platoon length be $L=10\text{m}$. Note that the platoon length cannot be smaller than $9$m. The projection of the safe set $\mathbb{S}$ onto the $\tilde x_1$-$\tilde x_2$ space is the following triangle: 
$$\left\{(\tilde x_1,\tilde x_2) \Big | \tilde x_1+4.5\le \tilde x_2, \tilde x_2 \le 10, \tilde x_1 \ge 4.5 \right \}.$$ We found $\lambda^*=0.23$ (with $0.01$ precision).
 The projection of $\Omega^*$ on $\tilde x_1$-$\tilde x_2$ is shown in Fig. \ref{fig:invariant}. 
\item Let $N=6$ and $L=30\text{m}$. We found $\lambda^*=0.29$. We simulated the system for 120 time steps, which equals 60 seconds. The initial conditions were the centers of the intervals for allowed spacings and speeds. The disturbances at each time were selected randomly from the set $\mathbb{W}$, with a bias toward the boundary of $\mathbb{W}$ (to simulate the system under heavy disturbances). We used $J(\mu)=\left\|\mu(y)\right\|_2^2$, which penalizes the total control effort. Thus, the control inputs at each time were computed using the following optimization problem:
%\vspace{-0.65em}
\begin{equation}
\begin{array}{lll}
u(y) & = \text{argmin} & \|u\|^2, \\
& s.t. & Ay +  Bu \oplus E\mathcal{W} \subseteq \Omega^*,
\end{array}
%\vspace{-0.15em}
\end{equation}  
which is mapped into a quadratic program and is feasible by construction for all $y \in \Omega^*$. 
The results in Fig. \ref{fig:invariant} show that the system specifications are met. The headways (the distances of the fronts of the vehicles from the backs of the vehicles immediately preceding them) are always greater than zero. Thus, collision avoidance is ensured. The platoon speed, length and controls are bounded within their specified ranges. The bottom plot in Fig. \ref{fig:invariant} shows the trajectory of the vehicles in the leader frame, where the vertical axis is time and the horizontal axis is space. Each vehicle is depicted by a black rectangle. The leader, which is the vehicle on the right, is fixed in this frame and the follower vehicles move due to the disturbances.
%\begin{figure}[t]
%\centering
%\vspace{8pt}
%%\includegraphics[width=0.23\textwidth]{headways}~
%%\includegraphics[width=0.23\textwidth]{speed}\\
%%\includegraphics[width=0.46\textwidth]{trajectories}
%%\includegraphics[width=0.23\textwidth]{r1}
%%\includegraphics[width=0.23\textwidth]{r2}\\
%%\includegraphics[width=0.23\textwidth]{r3}
%%\includegraphics[width=0.23\textwidth]{r4}
%\includegraphics[width=0.155\textwidth]{r2}
%\includegraphics[width=0.155\textwidth]{r3}
%\includegraphics[width=0.155\textwidth]{ex4}
%\\
%\vspace{3pt}
%\includegraphics[width=0.48\textwidth]{r0}
%\caption{Example 1: platoon of $N=6$. [Top, Clockwise]: Headway, platoon speed and controls [Bottom] Vehicular displacements in the leader frame.}
%\vspace{-0.5em}
%\label{fig:trajectories}
%\end{figure}
%\vspace{-0.000001em}
\item We study how $\lambda^*$ and the computation time of $\Omega^*$ vary with $N$. We keep the density $\rho=N/L$ constant and vary $N$. We used the Gurobi \cite{gurobi} \color{black} LP solver on a dual core 2.8GHz iMac. The results, shown in Table \ref{table:time}, indicate that $\lambda^*$ grows with $N$, even though $\rho$ is constant. For example, 4 vehicles following a leader within $20$m can accommodate larger disturbance magnitudes than 8 vehicles within $40$m. This is attributed to the advantage gained from coordinating the vehicles in a centralized manner. We also { empirically} observe that $\lambda^*$ converges to a particular value as $N$ increases. 
\begin{table}
\centering
\caption{$\lambda^*$ and computation times for $\Omega^*$ ({$\tau$ in seconds})}
\resizebox{0.49\textwidth}{!}{
\begin{tabular}{|c||c|c|c|c|c|c|c|c|}
\hline 
$N$ & $1$ & $2$ & $4$ & $6$ & $8$ & $10$ & $15$ & $20$ \\
\hline
$L$ & $5$ & $10$ & $20$ & $30$ & $40$ & $50$ & $75$ & $100$ \\
\hline
$\lambda^*$ & $0.17$ & $0.23$ & $0.28$ & $0.29$ & $0.31$ & $0.32$ & $0.33$ & $0.33$ \\
\hline
$\tau$ & $0.004$ & $0.028$ & $0.29$ & $1.36$ & $5.45$ & $37.8$ & $262$ & $2115$ \\
\hline
\end{tabular} 
}
%\vspace{-2em}
\label{table:time}
\end{table}
\end{enumerate} 
\end{example}
%\vspace{-1.5em}
\subsection{Distributed Control}
\label{sec:distributed}
In this section, we propose a distributed solution to Problem \ref{problem:problem}.  As discussed earlier, our approach requires V2V communication to measure relative distances and velocities. Note that distributed sensing is usually much cheaper than distributed decision making. We follow a divide and conquer approach to distributed control. We impose restrictions on the evolution of each vehicle such that control decisions can be made without a central coordination. The main advantage of our distributed framework is that the computation of control policies has $\mathcal{O}(1)$ complexity.

We make the following observations from inspecting \eqref{eq:relative_dynamics}. The state of each vehicle is not directly influenced by other vehicles (matrix $A$ is block-diagonal). However, the control decision and disturbances of the leader affects all the follower vehicles. Define $\tilde u_i:=u_0-u_i$ and $\tilde w_{s,i}:=w_{s,0}-w_{s,i}, s=x,v, i=1,\cdots,N$, as the new control input and disturbances,, respectively, of the follower vehicles. Also, define $\tilde u:=\{\tilde u_i\}_{i=0,1,\cdots,N}$, $\tilde w=\{\tilde w_{s,i}\}_{i=0,1,\cdots,N, s=x,v}$, and $\mathcal{U}$ and $\mathcal{W}$ as the admissible sets of $\tilde u$ and $\tilde w$, respectively. Note that $\tilde u_0=u_0, \tilde w_{0,s}=w_{0,s},s=x,v$. Now, $\mathcal{ U}$ and $\mathcal{ W}$ are no longer in hyperrectanglular form similar to \eqref{eq:u} and \eqref{eq:w}, but are still polyhedrons. 
We examine the relationship between $\mathbb{U}$ and $\mathcal{U}$ - the case of $\mathbb{W}$ and $\mathcal{ W}$ is similar but in higher dimensions. We have $\mathcal{ U}=T\mathbb{U}$, where
\begin{equation}
\label{eq:transformation}
T=
\left(
\begin{array}{cc}
1_N & -I_N  \\ 
%\hdashline
0\times 1^T_N & 1
\end{array}
\right).
\end{equation} 
A graphical representation of the transformation of a hyper-rectangle under the linear transformation $T$ is illustrated in Fig. \ref{fig:rectangle}. 
In order to obtain hyper-rectangular admissible sets, we under-approximate $\mathcal{ U}$ and over-approximate $\mathcal{ W}$ by hyper-rectangles $\underline{\mathcal{R}}_\mathbb{U}$ and $\overline{\mathcal{R}}_\mathbb{W}$, respectively (see Fig. \ref{fig:rectangle}). 
Over/under-approximating polyhedra by hyper-rectangles is conservative  but facilitates a decentralized architecture since hyper-rectangles are cartesian products of intervals. 
Therefore, we add conservatism but retain the hyper-rectangular structure of the set of admissible controls and disturbances. 
Basic properties of hyper-rectangles imply:
%\vspace{-0.35em}
\begin{gather*}
\tilde u_i \in Proj_{i}(\underline{\mathcal{R}}_\mathbb{U}), i=1,\cdots,N \Rightarrow \tilde u \in \mathcal{ U}. \\
\tilde w \in \mathcal{ W} \Rightarrow \tilde w_i \in Proj_{i,s}(\overline{\mathcal{R}}_\mathbb{W}), ~i=1,\cdots,N,~s=x,v.
%\vspace{-0.35em}
\end{gather*}
\begin{figure}[t]
\centering
\vspace{5pt}
\begin{tikzpicture}[xscale=1,yscale=1]
\draw[red,fill=red!50] (-0.5,0) -- ++(1,0) -- ++(0,1) -- ++(-1,0) -- cycle;
\draw[dashed,black,->] (0.0,0.5) -- ++(0.8,0);
\draw[dashed,black,->] (0.0,0.5) -- ++(0,0.8);
\node[] at (-0.3,1.2) {\small $ u_0$};
\node[] at (0.8,0.7) {\small $ u_1$};
\node[] at (-0.2,0.3) {\small $H$};

\draw[red,fill=red!50] (3.5,0.5) -- ++(0,1) -- ++(-1,0) -- ++(-1,-1) -- ++(0,-1) -- ++(1,0) --++(1,1) -- ++(0,1) -- cycle;
\node[] at (2.5,0.3) {\small $TH$};

\draw[dashed,black,->] (2.5,0.5) -- ++(0.8,0);
\draw[dashed,black,->] (2.5,0.5) -- ++(0,0.8);
\node[] at (2.3,1.1) {\small $\tilde u_0$};
\node[] at (3.2,0.7) {\small $\tilde u_1$};

\draw[blue,dashed,fill=cyan!80!white] (4.5,-0.5) -- ++(2,0) -- ++(0,2) -- ++(-2,0) -- ++(0,-2) -- cycle;
\draw[red,fill=red!50,opacity=0.5] (6.5,0.5) -- ++(0,1) -- ++(-1,0) -- ++(-1,-1) -- ++(0,-1) -- ++(1,0) --++(1,1) -- ++(0,1) -- cycle;
\draw[green,dashed,fill=green!50] (5,0) -- ++(1,0) -- ++(0,1) -- ++(-1,0) -- ++(0,-1) -- cycle;

\node[] at (5.5,0.5) {\small $\underline{\mathcal{R}}_H$};
\node[] at (4.8,1.2) {\small $\overline{\mathcal{R}}_H$};

\end{tikzpicture}
\caption{[Left]: hyper-rectangle $H$, [Middle]: Transformation of $H$ under the linear transformation \eqref{eq:transformation} ($N=1$), [Right]: The over-approximating hyper-rectangle $\overline{\mathcal{R}}_H$ (blue) and under-approximating $\underline{\mathcal{R}}_H$ (green). We have $\underline{\mathcal{R}}_H \subset H \subset \overline{\mathcal{R}}_H$.
}
%\vspace{-1.5em}
\label{fig:rectangle}
\end{figure}
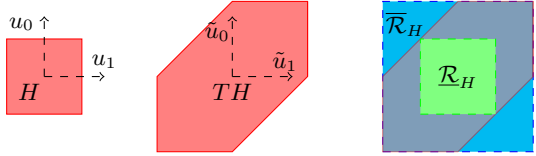
\begin{figure}[t]
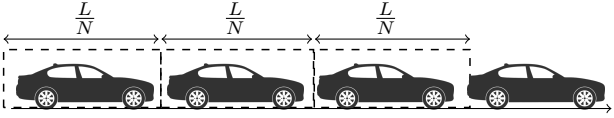

\centering
\vspace{-1pt}
\begin{tikzpicture}[xscale=2,yscale=0.7]
	\draw [dashed,line width=0.2mm] (  -0.54 , -0.4  ) rectangle (0.505 ,0.7 );
	\draw [dashed,line width=0.2mm] (  0.505 , -0.4  ) rectangle (1.520 , 0.7 );
	\draw [dashed,line width=0.2mm] (  1.525 , -0.4  ) rectangle (2.56 , 0.7);
    \node[] at (0,0) {\includegraphics[width=0.1\textwidth]{car}};
    \node[] at (1,0) {\includegraphics[width=0.1\textwidth]{car}};
    \node[] at (2,0) {\includegraphics[width=0.1\textwidth]{car}};
    \node[] at (3,0) {\includegraphics[width=0.1\textwidth]{car}};
    \draw[color=black,<-]  (3.5, -0.42 ) -- (-0.49, -0.42 );
    \draw[color=black,<->]  (-0.54, 0.9) -- (0.49, 0.9 );
    \draw[color=black,<->]  (0.51, 0.9) -- (1.51, 0.9 );
    \draw[color=black,<->]  (1.53, 0.9) -- (2.56, 0.9 );
    \node[] at (1,1.3) {$\frac{L}{N}$};
    \node[] at (2,1.3) {$\frac{L}{N}$};
    \node[] at (0,1.3) {$\frac{L}{N}$};
    \end{tikzpicture}
    %\vspace{-0.95em}
\caption{Envelopes for vehicle positions with respect to leader}
%\vspace{-1.85em}
\label{fig:envelopes}
\end{figure}
Next, we under-approximate the non-rectangular constraints in $\mathbb{S}$, i.e., the constraints in \eqref{eq:collide}, by rectangular constraints. For simplicity, we assume $l_i=l,i=0,1,\cdots,N$, for the rest of the paper. We define $\mathcal{S} \subset \mathbb{S}$ by replacing \eqref{eq:collide} with: 
{\small
	%\vspace{-0.35em}
\begin{equation}
\label{eq:envelope}
il+(i-1)\frac{L-Nl}{N} \le \tilde x_i \le il+ i\frac{L-Nl}{N}, i=1,\cdots,N.
\end{equation}}
The constraints in \eqref{eq:envelope} can be physically interpreted as \emph{envelopes} for each vehicle's position with respect to the leader (see Fig. \ref{fig:envelopes}). Every vehicle has to remain within its envelope for all times. Note that $\tilde x_N \le L$ follows from \eqref{eq:envelope}. From a divide and conquer point of view, every vehicle assumes that its neighbors do not enter its envelope (assumption), while every vehicle promises to not move beyond its envelope (guarantee). Therefore, correctness is guaranteed with no centralized coordination. Note that the computations of $\tilde u_i$'s are independent, but the computation of each $u_i,i=1,\cdots,N$ (actual control input) requires the control input of the leader, $u_i=u_0-\tilde u_i$. Hence every vehicle has to communicate with the leader. Having fully decoupled the dynamics and all constraints for each follower vehicle, we compute separate RCI sets $\Omega_i, i=1,\cdots, N$, where 
$\Omega_i \subseteq \mathbb{S}_i$, $\mathbb{S}_i=\Big \{ (\tilde x_i,\tilde v_i) \big | il+(i-1)\frac{L-Nl}{N} \le \tilde x_i \le il+ i\frac{L-Nl}{N} \Big\}$, and we have
%\vspace{-0.15em}
\begin{equation*}
\begin{array}{l}
\forall (\tilde x_i,\tilde v_i) \in \Omega_i, \exists \tilde u_i \in Proj_{i}(\underline{\mathcal{R}}_\mathbb{U}),
\\
 ( \tilde x_i +\tilde v_i \Delta \tau + \tilde u_i \frac{\Delta \tau^2}{2} + \tilde w_{x,i}, \tilde v_i + \tilde u_i \Delta \tau + w_{i,v}) \in \Omega_i, 
\\
\forall w_{i,x} \in Proj_{i,x}(\overline{\mathcal{R}}_\mathbb{W}),
 \forall w_{i,v} \in Proj_{i,v}(\overline{\mathcal{R}}_\mathbb{W}).
\end{array}
\end{equation*} 
We then define the invariance inducing control policy
$\mu_i:\Omega_i \rightarrow Proj_{i}(\underline{\mathcal{R}}_\mathbb{U}), u=1,\cdots,N.$
Finding the RCI set $\Omega_0 \subseteq \mathbb{S}_0$ for the leader, where $\mathbb{S}_0= [v_{0,\min},v_{0,\max}]$, we have:
%\vspace{-0.35em}
\begin{equation*}
\begin{array}{l}
\forall v_0 \in \Omega_0, \exists u_0 \in Proj_{0}(\underline{\mathcal{R}}_\mathbb{U}),
\\ v_0+u_0 \Delta \tau + [w_{0,v,\min},w_{0,v,\max}] \in \Omega_0.  
\end{array}
%\vspace{-0.35em}
\end{equation*} 
Similarly, we define the invariance inducing control policy 
$\mu_0:\Omega_0 \rightarrow Proj_{0}(\underline{\mathcal{R}}_\mathbb{U}).$

\begin{proposition}
We have $\mathcal{S} \subset \mathbb{S}$, where $\mathcal{S}=\prod_{i=0}^N \mathbb{S}_i$, and $\prod_{i=0}^N \Omega_i$ is a RCI set in $\mathcal{S}$. 
\end{proposition} 
\begin{proposition}
We have $<\mu_i>_{i=0,1,\cdots,N} \in \mathbb{M}$.
\end{proposition} 

\begin{IEEEproof}(sketch)
Correctness follows from the fact that we have under-approximated the admissible control inputs and over-approximated the admissible disturbances. 
\end{IEEEproof}

%\vspace{-0.5em}
\begin{example}
	\begin{figure}[t]
		%\vspace{5.5pt}
		\centering
		\includegraphics[width=0.25\textwidth]{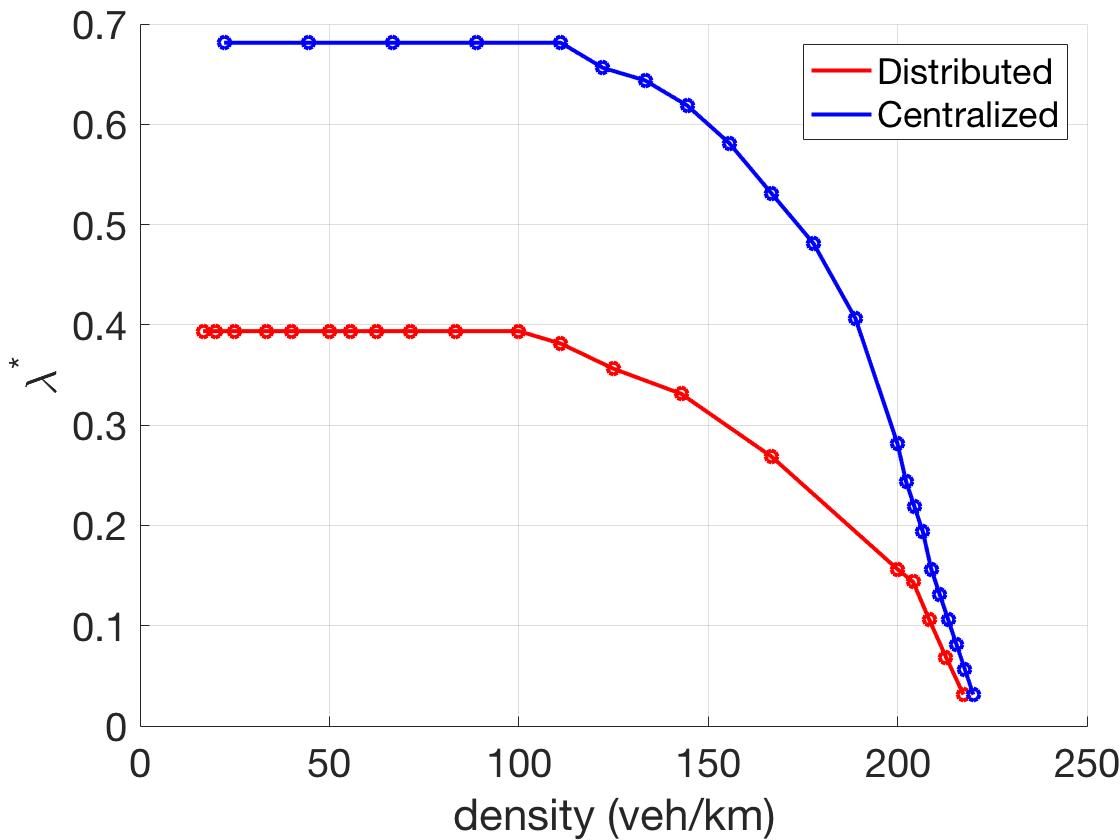}
		\vspace{-0.5em}
		\caption{Example 2: Density vs the magnitude of disturbances that the platoon can accommodate for centralized and distributed control architecture.}
		%\vspace{-1.5em}
		\label{fig:controllers}
	\end{figure}
Consider the specification in Example \ref{example:centralized}. We implement the procedure explained in this section to design a distributed control policy. For every vehicle $i=0,1,\cdots,N$, the admissible set for $\tilde u_i$ is  $[-3/2,3/2] \text{m/s}^2$ 
- ranges halved as compared to $\mathbb{U}$ in Example \ref{example:centralized} - and the admissible sets for $\tilde w_{i,x}$ and $\tilde w_{i,v}$ are $[-2,2]\text{m/s}$ and $[0.5,0.5]\text{m}$, respectively - ranges doubled as compared to $\mathbb{W}_0$ in Example \ref{example:centralized}. The computation time is negligible as the dimension of each RCI set is 2 (1 for the leader).  We vary the density $N/L$, find the corresponding $\lambda^*$ and compare the result to the one obtained using the centralized control policy (Fig. \ref{fig:controllers}). It is observed that the distributed control architecture can accommodate a smaller magnitude of disturbances. We expect the ranges in $\mathbb{W}$ to have an increasing relationship with average speed of the platoon. Therefore, Fig. \ref{fig:controllers} suggests that platoons that move faster should have wider inter-vehicular spacings to be considered safe.
\end{example}

We end this section with a final remark. Since the control inputs and disturbances of the leader affect all the vehicles in the platoon, a perfect leader with zero disturbances (and subsequently controls, as explained shortly) will eliminate the dynamical coupling between the vehicles in the platoon. This observation leads to the following statement.  
\begin{proposition}[Perfect Leader]
Consider a platoon with identical follower vehicles and disturbances as described in Section \ref{sec:problem}. If there are no disturbances acting on the leader, that is, $w_{0,x}=w_{0,y}=0$, then the $\lambda^*$s for the centralized and distributed control policies are equal. 
\end{proposition} 
\begin{IEEEproof}
(sketch) If there are no disturbances acting on the leader, then $u_0=0$ is a valid control policy that maintains the platoon speed within the desired range, provided that it is initially within the range. With $w_{0,x}=w_{0,y}=u_0=0$, the only coupling between the vehicles is through the collision constraints in \eqref{eq:collide}, which are replaced by \eqref{eq:envelope} in the distributed architecture. We need to show that centralized controllers cannot handle larger $\lambda^*$ by violating \eqref{eq:envelope}. Suppose a valid centralized control policy violating \eqref{eq:envelope}. It follows from \eqref{eq:length} and Fig. \ref{fig:envelopes} that at least one vehicle (e.g., the last vehicle) has to remain in its envelope when other vehicles move out of theirs. As the dynamics are independent, there exists a control policy to keep the (last) vehicle inside its envelope. Since the vehicles are identical and the disturbances are independent of each other, such a control policy should be valid to keep any vehicle in its envelope. Therefore, the existence of any centralized policy implies the existence of distributed control policies. Hence, a distributed control policy must exist for the same $\lambda^*$. 
\end{IEEEproof}

\section{Conclusion and Future Work}
We utilized results in set-invariance to formally synthesize platoon cruise control policies subject to collision avoidance requirements. We found a centralized policy as well as a computationally attractive (but conservative) distributed policy. {We will consider more relaxed platoon information structures such as predecessor-following in future work.} 

%\balance

\bibliography{references2}  

\balance
%%%%%%%%%%%%%%%%%%%%%%%%%%%%%%%%%%%%%%%%%%%%%%%%%%%%%%%%%%%%%%%%%%%%%%%%%%%%%%%

\end{document}